\documentclass[conference]{IEEEtran}
\IEEEoverridecommandlockouts

\usepackage{graphicx}
\usepackage{booktabs}
\usepackage{xcolor}

\usepackage{amsmath}
\usepackage{amssymb}
\usepackage{cite}
\usepackage{cuted}
\newcommand{\linebreakand}{%
  \end{@IEEEauthorhalign}
  \hfill\mbox{}\par
  \mbox{}\hfill\begin{@IEEEauthorhalign}
}    
\usepackage{bbm}
\usepackage{algorithm} 
\usepackage{algpseudocode}
\usepackage{caption}
\usepackage{steinmetz}
\usepackage{subcaption}
\usepackage{amsthm}
\usepackage[english]{babel}


\newcounter{relctr} 
\everydisplay\expandafter{\the\everydisplay\setcounter{relctr}{0}} 
\usepackage[figurename=Fig.]{caption}

\usepackage{xpatch}

\newtheoremstyle{remarkstyle}%
  {}
  {}
  {\itshape}
  {}
  {\itshape}
  {.}
  {.5em}
  {}

\theoremstyle{remarkstyle}

\AtBeginDocument{} 

\ifCLASSINFOpdf
\else
\fi

\begin{document}

\title{Sequential Monte Carlo for Network Resilience Assessment and Control
\thanks{This work is partially supported in Finland by the Research Council of Finland (Grants 362782 and 369116 (6G Flagship)).}
}

\author{
Onel~L.~A.~L\'{o}pez, \emph{IEEE Senior Member}\\
University of Oulu, Finland, onel.alcarazlopez@oulu.fi}



\maketitle

\begin{abstract}

Resilience is emerging as a key requirement for next-generation wireless communication systems, requiring the ability to assess and control rare, path-dependent failure events arising from sequential degradation and delayed recovery. In this work, we develop a sequential Monte Carlo (SMC) framework for resilience assessment and control in networked systems. Resilience failures are formulated as staged, path-dependent events and represented through a reaction-coordinate-based decomposition that captures the progression toward non-recovery. Building on this structure, we propose a multilevel splitting approach with fixed, semantically interpretable levels and a budget-adaptive population control mechanism that dynamically allocates computational effort under a fixed total simulation cost. The framework is further extended to incorporate mitigation policies by leveraging SMC checkpoints for policy evaluation, comparison, and state-contingent selection via simulation-based lookahead. A delay-critical wireless network use case is considered to demonstrate the approach. Numerical results show that the proposed SMC method significantly outperforms standard Monte Carlo in estimating rare non-recovery probabilities and enables effective policy-driven recovery under varying system conditions. The results highlight the potential of SMC as a practical tool for resilience-oriented analysis and control in future communication systems.

\end{abstract}


\IEEEpeerreviewmaketitle

\section{Introduction}

\IEEEPARstart{S}{ixth}-generation (6G) wireless systems may consider resilience as a first-class requirement \cite{Alves.2025}. Unlike current network designs, which emphasize good performance and fault-avoidance mechanisms under nominal operating conditions, resilience encompasses a system's ability to withstand, adapt to, and recover from complex, dynamic, and potentially cascading disruptions \cite{Rak.2020,Matthiesen.2025}. Indeed, future networks must operate as critical infrastructures, tightly coupled with other societal systems, where failures may propagate across domains and time scales, and must incorporate capabilities such as graceful degradation, rapid reconfiguration, and learning-driven recovery \cite{Alves.2025}. These capabilities naturally define a sequence of system operating stages, from nominal operation through degradation and critical regimes to eventual failure and recovery, which must be explicitly accounted for to enable rigorous analysis and control.

Assessing and engineering resilience in 6G systems requires advanced statistical tools capable of capturing rare but impactful events, particularly those arising from sequential degradation or cascading failures \cite{Alves.2025}. Simulation plays a central role in this process, enabling the exploration of complex system dynamics and the anticipation of failure regimes before they occur in practice.\footnote{A related paradigm is that of digital twins, which are inherently data-driven (i.e., periodically updated through measurements from the physical system). Note that the collected data can be leveraged to learn stochastic models for model-based simulation and sampling of system dynamics.} However, classical Monte Carlo (MC) becomes prohibitively inefficient in this setting \cite{Rubino.2009,Lopez.2023}, especially when failures emerge through progressive transitions across intermediate regimes rather than isolated events. 
Indeed, efficient estimation requires methodologies that explicitly exploit the sequential structure of the progression toward failure, enabling targeted exploration of critical system trajectories.

Sequential MC (SMC) constitutes a powerful framework for addressing these challenges by constructing intermediate probability distributions and approximating them using interacting particle systems, where trajectories are replicated as they approach increasingly critical regions of the state space \cite{Moral.2006,Rubino.2009,Smith.2013}. There are different SMC methods that leverage, e.g., importance splitting \cite{Villen.1991,Garvels.2000,Rubino.2009,Garvels.2002}, subset simulation \cite{Lopez.2023,Bect.2017}, and adaptive multilevel splitting \cite{Cerou.2007}. They all build on the fact that rare failures are a consequence of progressive transitions across levels and demonstrate substantial efficiency gains over standard MC. These aspects certainly make SMC particularly well-suited for systems exhibiting staged degradation and resilience dynamics.


Despite SMC's maturity, its application to network resilience studies remains largely limited, with existing works focused on reliability/robustness assessment, e.g., \cite{Budde.2020,Lopez.2023}.\footnote{Notably, most rare-event simulation frameworks for wireless networks use non-sequential variants instead of SMC, e.g., \cite{Yu.2020,Ke.2023,Yu.2025}. Such approaches, although suitable for ultra-reliable low-latency communication systems, which is indeed their focus, do not fit well with the time/stage dynamics that must be accounted for in network resilience studies.} 
However, resilience involves a broader and more dynamic perspective, encompassing not only failure occurrences but also the evolution of system states across multiple regimes, including degradation, adaptation, and recovery. This highlights a gap between classical rare-event simulation methodologies and the requirements of resilience-aware system design. We address this gap in this work by showcasing SMC for resilience assessment and control in networked systems. Specifically, we i) formulate resilience in terms of staged system dynamics and introduce a reaction-coordinate-based representation that captures the progression toward failure; ii) propose an SMC methodology that exploits this structure through multilevel splitting with adaptive population control under a fixed computational budget, enabling efficient estimation of rare resilience-related events; iii) extend the framework to incorporate control actions, enabling policy selection based on their impact on future system evolution; and iv) demonstrate the effectiveness of the proposed approach through a representative delay-critical wireless network use case, illustrating its ability to efficiently capture rare failure and non-recovery probabilities and inform resilience-oriented decision-making.
\vspace{-2mm}
\section{Preliminaries of SMC}\label{sec:background}
%
%
Consider a stochastic dynamical system $X(t)\in\mathcal{X}, t\ge 0$, and a \emph{fault (or failure) set} $\mathcal{F}\subset \mathcal{X}$, representing undesirable system configurations. We define the path-dependent event
\begin{align}
    \xi\triangleq \{\exists t\le T: X(t)\in\mathcal{F}\},\label{xi}
\end{align}
where $T$ is an observation horizon. Then, $p\triangleq \Pr(\xi)$ quantifies the likelihood of entering an undesirable regime within $T$, and we focus on its estimation. For this, assume access to a stochastic simulator that generates sample paths $X^{(i)}(t)$.
%
%
\vspace{-2mm}
\subsection{N\"aive MC}
\vspace{-1mm}
A n\"aive MC approach estimates $p$ using $N$ independent full-trajectory realizations of the stochastic process 
as
%
\begin{align}
    \hat{p}_{\text{mc}}=\frac{1}{N}\sum_{i=1}^N\mathbf{1}(\xi^{(i)}),
\end{align}
where $\xi^{(i)}\triangleq \{\exists t\le T: X^{(i)}(t)\in \mathcal{F}\}$ similar to \eqref{xi}, and the indicator equals one if the $i$-th trajectory enters the fault set at any time prior to $T$. This estimator is unbiased, i.e., $\mathbb{E}[\hat{p}_{\text{mc}}]=p$, with variance $\text{Var}[\hat{p}_{\text{mc}}]=p(1-p)/N$, implying a sample complexity $\mathcal{O}(1/p)$ in the rare-event regime \cite{Lopez.2023}. This can be easily seen from 
the relative variance
\begin{align}
   \text{Var}_r[\hat{p}_{\text{mc}}] =\frac{\text{Var}[\hat{p}_{\text{mc}}]}{\mathbb{E}[\hat{p}_{\text{mc}}]^2}=\frac{1-p}{p N},
\end{align}
since $\text{Var}_r[\hat{p}_{\text{mc}}]\rightarrow 1/(p N)$ as $p\rightarrow 0$. The method therefore struggles in the rare-event regime, e.g., $p\le 10^{-6}$, as its complexity explodes. 
As a measure of overall computational costs, we adopt a budget-based metric equal to the total number of base-simulator time steps executed, i.e.,
\begin{align}
    C_\text{mc}=\sum_{i=1}^N Q_i,
\end{align}
where $Q_i$ is the $i-$th path length. For rare events, most trajectories run until $T$, yielding $C_\text{mc}\approx N T/\Delta$.
\vspace{-2mm}
\subsection{SMC}\label{sec:smc}
\vspace{-1mm}
SMC addresses rare-event estimation by sequentially decomposing $\xi$ into a sequence of less rare conditional events, thereby avoiding the need to observe $\xi$ directly \cite{Cerou.2007,Lopez.2023}. 

Let us introduce a reaction coordinate $g: \mathcal{X}\rightarrow \mathbb{R}$, which quantifies the system's progression toward $\mathcal{F}$. Now, define an increasing sequence of thresholds (levels) $\ell_0 < \cdots<\ell_K$, with associated nested sets
%
    $\mathcal{L}_k \triangleq \{x\in \mathcal{X}: g(x)\ge \ell_k \}$,
such that $\ell_0=-\infty$, hence $\mathcal{L}_0=\mathcal{X}$, and $\mathcal{L}_K\subseteq \mathcal{F}$. The rare event probability can then be factorized as
\begin{align}
    \Pr(\xi)= \prod_{k=0}^{K-1}p_k,\label{pre}
\end{align}
where $p_k\triangleq \Pr(\xi_{k+1}|\xi_k)$, 
$\xi_k\triangleq \{\exists t\le T: X(t)\in \mathcal{L}_k\}$ (and note that $p_0=\Pr(\xi_1)$). The aim is to estimate each factor $p_k$, and then recover $\Pr(\xi)$ via multiplication.

SMC requires simulators to be restartable with independent continuation \cite{Villen.1991,Garvels.2000,Moral.2006}, i.e., able to resume from an intermediate state using fresh randomness. This holds for most event-driven and time-stepped simulators.
%

There are several SMC variants in the literature, differing in how levels are selected and how trajectories are replicated \cite{Villen.1991,Garvels.2000,Garvels.2000,Moral.2006,Rubino.2009,Smith.2013,Cerou.2007}. 
Herein, we focus only on a fixed-level splitting approach, not only for simplicity but also because it is appealing for resilience studies, as discussed later in Section~\ref{sec:Res}. 

\subsubsection{SMC pseudocode}
Let $M$ denote the number of particles (trajectories) maintained at each level and $\xi_k^{(i)}\triangleq \{\exists t\le T: X^{(i)}(t)\in \mathcal{L}_k\}$. The algorithm is as follows:
\begin{itemize}
    \item \textbf{initialization:} generate $M$ 
    i.i.d. initial simulator states $X^{(i)}(0)$ from the nominal initial-state distribution (or a deterministic initial condition);
    \item \textbf{level-by-level evolution:} for level $k=0,\cdots,K-1$, simulate each particle $i$ until it either reaches $\mathcal{L}_{k+1}$ (i.e., $\xi_k^{(i)}$ is non-empty) or $T$ (i.e., $\xi_k^{(i)}$ is empty). Upon hitting $\mathcal{L}_{k+1}$, store a checkpoint including all simulator state required for restarting in the subsequent splitting step;
    \item \textbf{selection and splitting:} let $S_{k}$ be the number of trajectories that reach $\mathcal{L}_{k+1}$ at level $k$. 
    Retain them and resample with replacement from their checkpoints to restore the population to $M$. 
    Each resampled checkpoint initializes a new simulator instance, which is continued independently using fresh random input realizations.
    \item \textbf{estimation:} $\hat{p}_k=S_{k}/M$, and use \eqref{pre} such that
    %
        $\hat{p}_{\text{smc}}=\prod_{k=0}^{K-1}\frac{S_{k}}{M}$,\label{smc}
    which is unbiased under ideal i.i.d. sampling.
\end{itemize}

\subsubsection{Accuracy and cost} The estimation variance depends primarily on the variance of the stage-wise conditional probabilities $p_k$. In fact, for moderate $\{p_k\}$ and  sufficiently large $M$, a standard, engineering approximation (ignoring inter-stage dependence due to resampling) yields \cite{Bect.2017}
\begin{align}
    \text{Var}_r[\hat{p}_{\text{smc}}] \approx \sum_{k=0}^{K-1}\frac{1-p_k}{p_k M}.\label{eq:var}
\end{align}
Meanwhile, the computational cost per level is $C_k=\sum_{i=1}^{M}Q_{k,i}$, where $Q_{i,k}$ is the number of steps the particle $i$ is propagated while attempting to go from level $k$ to $k+1$. The total cost is then given by
\begin{align}
    C_\text{smc}=\sum_{k=0}^{K-1}C_k. 
    \label{eq:Csmc}
\end{align}
The levels must be chosen such that $\{p_k\}$ are moderate, as small $p_k$ leads to high variance, whereas large $p_k$ requires many levels, increasing cost and reducing interpretability.
\vspace{-1mm}
\section{SMC for Resilience Assessment \& Control}\label{sec:Res}
\vspace{-1mm}
 A ``resilience'' failure event can be defined using \eqref{xi} by making $\mathcal{F}$ to encode persistence and/or recovery-deadline constraints.
Since such events are rare and path-dependent, splitting-based techniques that exploit progressive degradation are particularly suitable.
Herein, we discuss how the basic framework from Section~\ref{sec:smc} can be exploited for this.
\subsection{Simulation Design}
Key elements in SMC design/implementation are the definition of $g(\cdot)$, selection of hitting levels \cite{Garvels.2002,Cerou.2007}, and population control, as we discuss in the following. 
\subsubsection{Reaction coordinate} $g(\cdot)$ quantifies the system's progression toward the rare event set. 
In resilience assessment, this choice is particularly critical, as it determines not only estimator efficiency but also the interpretability of intermediate system states.
Such a function must:
%
    i) increase toward failure in expectation \cite{Garvels.2002,Cerou.2007}, and ii)
    remain operationally interpretable, such that the coordinate corresponds to meaningful service/control-relevant figures.

Note that $g(\cdot)$ is evaluated on the instantaneous simulator state $X(t)$, possibly including augmented state variables that encode temporal context. 
Natural reaction coordinate candidates in communication networks include backlog/delay levels,  accumulated stress or impairment measures, and recovery slack variables that quantify the remaining time before a recovery deadline is violated. Hybrid coordinates that combine state and temporal information are particularly useful when resilience violations are defined through persistence conditions rather than instantaneous thresholds.

\subsubsection{Hitting levels}
The adopted fixed-level splitting in Section~\ref{sec:smc} can promote interpretability
by aligning levels with resilience phases (e.g., nominal operation, degradation, SLA violation, and non-recovery), while the corresponding $p_k$ quantifies stage-wise vulnerability.
In general, the choice of level spacing/sets involves a trade-off between statistical efficiency and interpretability. This is because finely spaced levels reduce estimator variance but increase algorithmic complexity and may obscure the physical meaning of intermediate transitions, while overly coarse levels improve clarity but can lead to high variance or particle extinction if conditional transition probabilities are too small \cite{Garvels.2002,Cerou.2007}. In resilience-oriented studies, moderate conditional probabilities combined with meaningful phase boundaries may be preferable to aggressive variance minimization.

\subsubsection{Population control}\label{sec:PC}
Because the levels $\{\ell_k\}$ are defined a priori based on semantic notions of service degradation and persistence, the conditional probabilities $\{p_k\}$ may differ significantly. 
This can lead to particle extinction at difficult levels or inefficient computational effort allocation.
To address this, we adopt a budget-adaptive sampling strategy in which the total available computational cost $C_T$ is fixed a priori, and simulation effort is dynamically allocated across levels. Specifically, trajectory continuations are generated sequentially at each level $\mathcal{L}_k$ from the current state pool until both the number of trajectories $S_k$ that successfully reach the next level $\mathcal{L}_{k+1}$ and the total number of attempts $A_k$ are collected. That is, until conditions $S_k\ge S_{\text{tar}}$ and $A_k\ge A_{\text{tar}}$ are simultaneously met or the global budget is exhausted earlier.
After each level, successful trajectories are resampled to form the input pool for the next level, with the pool size dynamically adjusted to be inversely proportional to $S_\text{tar}/\hat{p}_k$, where $\hat{p}_k=S_k/A_k$.
The resulting estimator retains the multiplicative structure in \eqref{pre}, becoming
\begin{align}
    \hat{p}_{\text{smc}}=\prod_{k=0}^{K-1}\frac{S_k}{A_k},
\end{align}
while introducing only a minor bias due to outcome-dependent stopping (see discussions in \cite{Moral.2006,Bect.2017}), which can be controlled through relatively large values of $S_{\text{tar}}$ and $A_{\text{tar}}$. Note that the computational cost is still given by \eqref{eq:Csmc} but now with $C_k=\sum_{i=1}^{A_k}Q_{k,i}$.
%

This approach relates to other advanced SMC methods that also rely on resampling and dynamic allocation of computational effort, e.g., \cite{Moral.2006,Smith.2013,Bect.2017,Cerou.2007}. However, instead of adapting the levels, our proposed formulation explicitly constrains the total computational budget and distributes it across levels through data-driven stopping criteria. This provides a simple and practical mechanism to concentrate simulation effort on the most critical transitions without requiring prior tuning of population sizes. Also, and more importantly, it retains the interpretability of fixed-level splitting, making it particularly suitable for resilience analysis.
\vspace{-1mm}
\subsection{Mitigation Policies}\label{sec:reconf}
\vspace{-1mm}
SMC also offers potential for fault mitigation and control policies, crucial for resilient systems. Note that policies are causal decision rules that modify system dynamics, affecting disturbance, recovery dynamics, and/or corrective actions. The key idea here is to reuse the near-critical system states generated during splitting as representative starting points for controlled interventions via policies.

Let $u\in\mathcal{U}$ denote a fixed policy that may encompass preventive, reactive, or recovery-oriented actions. Under policy $u$, the system state evolves according to a controlled stochastic process $X^u(t)$, 
inducing a policy-dependent resilience failure probability
%
    $p(u)\triangleq \Pr(\xi|u)$. 
%
Equivalently, policies modify the simulator update rule, resulting in policy-dependent state transitions while preserving restartability.
The level sets remain policy-invariant to preserve unbiasedness, comparability of $p(u)$ across policies, and attribution of performance differences to policy effects. 

Given the above, the conditional probabilities
%
    $p_k(u)\triangleq \Pr(\xi_{k+1}|\xi_k,u)$
%
quantify how a given policy influences the likelihood of transitioning between successive resilience phases, hence providing fine-grained insight into where and how a policy is effective. 
This quantity is estimated via MC by branching each stored checkpoint at level $\mathcal{L}_k$ into multiple continuations under policy $u$ to the next level $\mathcal{L}_{k+1}$. These checkpoints correspond to simulator states at the first hitting time of $\mathcal{L}_k$ and thus represent samples from the conditional (first-hitting) distribution induced by the baseline dynamics.

\subsubsection{Comparing Policies} Policy comparison proceeds by:
%
i) running baseline SMC and storing checkpoints, ii) branching each checkpoint under candidate policies, iii) estimating $p_k(u)$, and iv) comparing conditional failure probabilities per policy, as well as auxiliary performance and cost metrics, when available.
    as well as auxiliary performance and cost metrics, when available.
%
Note that reductions in early-stage $p_k$ indicate improved robustness, while reductions in later stages reflect enhanced recovery or containment.

\subsubsection{Online mitigation/control}
\begin{figure}[t!]
    \centering
    \includegraphics[width=1\linewidth]{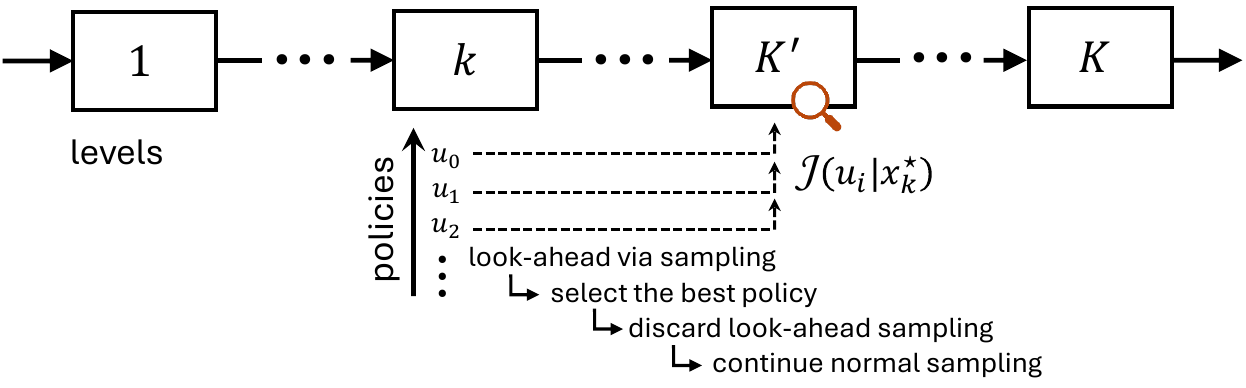}
    \caption{Illustration of the online mitigation/control procedure.}
    \label{fig:lookahead}
    \vspace{-4mm}
\end{figure}

Beyond policy evaluation and comparison, the framework can be extended to state-contingent policy selection at predefined resilience levels as illustrated in Fig.~\ref{fig:lookahead}. Specifically, when a trajectory first reaches a level policy-hosted $\mathcal{L}_k$ at state $x_k^\star$, policy selection is performed via simulation-based lookahead, evaluating candidate policies $\{u_i\}$ through inner simulations and selecting
\begin{align}
    u^\star(x_k^\star)=\arg\min_{u_i\in\mathcal{U}} \mathcal{J}(u_i|x_k^\star)
\end{align}
according to a predefined decision criterion $\mathcal{J}(\cdot)$.

An intuitive model for $\mathcal{J}$ is
\begin{align}
    \mathcal{J}(u_i|x_k^\star)=\sum_{l=k}^{K'}\ln \hat{p}_l(u_i) + c_k(u_i), \label{eq:LJ}
\end{align}
where $c_k(\cdot)$ captures the system cost (according to some metric) of implementing policy $u_i$. The impact of the conditional probabilities is linearized here to align with the behavior of common cost performance metrics. Meanwhile, $K'$ denotes a lookahead level such that $k\le K'\le K-1$. For instance, in a myopic setting, one has $K' = k$ such that \eqref{eq:LJ} leads to
%
the policy that just reduces the chances of progressing one level in the fault chain with reasonable cost. 

%

\section{A Use Case Example}\label{sec:system}

We showcase the proposed framework in a delay-critical wireless network system under stress and recovery dynamics. 

\subsection{System Model}\label{sec:model}
The system is observed over a finite horizon $T$, discretized into time steps of duration $\Delta=T/J$ and indexed by $j=0,\cdots,J$. We consider a service queue fed by a constant normalized arrival workload $\Lambda$,
expressed relative to nominal capacity. The normalized instantaneous service capacity at time step $j$ is denoted by $C[j]\in(0,1)$,
while $B[j]$ is the corresponding backlog, representing the amount of unfinished work measured in equivalent service time and evolving as
%
\begin{align}
    B[j+1]=\max\{0, B[j]+(\Lambda-C[j])\Delta\}.\label{eq:B}
\end{align}


The service capacity dynamics are driven by environmental stress and accumulated degradation, with a logistic mapping capturing saturation effects that are common in nonlinear systems, e.g., \cite{Akhter.2023}. Specifically, we adopt
\begin{align}
    C[j]&\triangleq (1+e^{-\eta[j]})^{-1}, \label{eq:C} 
\end{align}
where $\eta[j]\in\mathbb{R}$ is a latent health state given by
\begin{align}
    \eta[j+1]&=\eta[j]+\nu(u)(1-C[j])^{\varphi(u)} - F'[j].\label{eq:eta}
\end{align}
Here, $\nu(u)>0$ is the recovery rate and $\varphi(u)>1$ controls the nonlinearity of the recovery dynamics, both controlled by a recovery acceleration action $u$.
The recovery rate models adaptive system response that increases when the system operates below capacity, while the exponential term captures nonlinear amplification of stress effects, consistent with models of cascading failures and overload phenomena in complex networks \cite{Rak.2020}. Meanwhile, $F'[j]\triangleq e^{F[j]}$ comprises log-normal fatigue effects using a standard autoregressive process to capture temporal correlation in environmental disturbances as
\begin{align}
    F[j+1] = \rho F[j] + (1-\rho)\mu_F + \gamma\sigma_F, \label{eq:relax}
\end{align}
where $\rho\in[0,1)$ controls temporal correlation, $\mu_F\in\mathbb{R}$ is the long-term mean of the latent stress level with $\sigma_F> 0$ determining its variability, and $\gamma\sim\mathcal{N}(0,1)$. 

 The end-to-end delay experienced by traffic is approximated using Little's law as 
\begin{align}
    D[j]=B[j]/C[j]. \label{eq:delay}
\end{align} 
 %
We assess resilience through the system's ability to recover from a (rare) significant delay increase within a prescribed time window. For this, let
%
    $j_c\triangleq \inf\{j: D[j]\ge \delta\}$
denote the first time index at which the critical service delay threshold $\delta$ is exceeded. Once this event occurs, a recovery deadline of duration $t_\text{tar}$ is imposed. 
Specifically, a resilience break occurs if, after the first critical affectation, the delay remains above the critical threshold continuously throughout the grace period. This is mathematically represented by
\begin{align}
    \xi \triangleq \{\exists j_c\le J-H: D[j]\ge \delta, \forall j\in[j_c,j_c+H]\}, \label{eq:xi2}
\end{align}
where $H\triangleq \lceil t_\text{tar}/\Delta\rceil$ is the grace period expressed in discrete-time steps. 
This captures a path-dependent failure event.



\subsection{Resilience Failure Probability Estimation}\label{sec:resR}
Let the system state at time step $j$ be defined as 
\begin{align}
    X[j]\triangleq (B[j],C[j], F[j], \varrho[j]),\label{eq:X}
\end{align}
where $B[j]$, $C[j]$, and $F[j]$ evolve according to \eqref{eq:B}$-$\eqref{eq:relax}, while 
$\varrho[j]\in\{0,1,\cdots,H\}$ is a persistence variable counting the number of consecutive time steps for which the delay has remained above the critical threshold $\delta$. The latter evolves according to the recursion
%
\begin{align}
   \varrho[j+1]\!=\!\left\{\begin{array}{ll}
        \min(\varrho[j]+1,H), & D[j]\ge \delta  \\
        0, & D[j]<\delta
   \end{array}\right.\!\!,\ \forall j<J,\label{eq:z}
\end{align}
with the initial condition $\varrho[0]=0$, and where $D[j]$ is given by \eqref{eq:delay}.
This ensures that i) $\varrho[j]=d$ iff the delay has exceeded $\delta$ for $d$ consecutive steps, and ii) $\{X[j]\}$ is a time-homogeneous Markov process.
With this, the resilience failure event $\xi$ in \eqref{eq:xi2} is equivalently the hitting of a failure set $\mathcal{F}=\{(B,C,\varrho): \varrho=H\}$ as
%
    $\xi\triangleq \{\exists j: \varrho[j]=H\}$, 
%
and the goal is to estimate $\Pr(\xi)$.

A meaningful choice for the reaction coordinate is
\begin{align}
    g(X[j])=\min\Big(\frac{D[j]}{\delta},1\Big) + \frac{\varrho[j]}{H}\ \in[0,2],
\end{align}
as it captures the end-to-end delay up to the critical deadline $\delta$, shifting then the focus to the recovery delay. Then, one has 
%
    $0=\ell_0<\cdots<\ell_K=2$, and we use
%
$K=4$ with $\ell_1=1/10$, $\ell_2=1$, and $\ell_3=3/2$, capturing progression towards critical service affectation, the start of a critical service affectation, and mid-progression towards non-recovery. 

\begin{table}[t]
\centering
\caption{Baseline parameters for simulation.}
\label{tab:baseline_params}
\begin{tabular}{lll}
\toprule
\textbf{parameter} & \textbf{description}& \textbf{value / range} \\
\midrule
$\Delta$ & discrete-time step & $50$ ms \\
$T$ & simulation horizon & $60$ s \\
$B[0]$ & initial backlog condition & $0$\\
$\Lambda$ & offered traffic rate & 0.7 \\
$\eta[0]$ & initial/nominal recovery rate & $0.95$\\
$\nu(u)$ & recovery (relaxation) rate & $0.2$ \\
$\varphi(u)$ & non-linearity recovery parameter & 2\\
$\rho$ & temporal correlation of the latent stress & 0.75\\
$\mu_F$ & long-term mean of the latent stress  & -5 \\
$\sigma_F$ & standard deviation of the latent stress  & $0.55$ \\
$\delta$ & service degradation threshold & 100 ms \\
$t_\text{tar}$ & recovery target time & 5 s\\
$K$ & number of SMC levels & 4\\
$\ell_0,\cdots,\ell_4$ & SMC levels' thresholds & $0,0.1,1,1.5,2$\\
$C_T$ & total simulation cost & $5\times 10^6$\\
\bottomrule
\end{tabular}
\end{table}

\subsubsection{SMC procedure}
We adopt the fixed-level splitting procedure from Section~\ref{sec:smc}, incorporating the population-control mechanism from Section~\ref{sec:PC} to ensure numerical stability across levels.
Each particle/trajectory corresponds to a discrete-time realization of the system dynamics.  
When a particle reaches a new level, its state and time index are stored as a checkpoint and replicated, allowing multiple independent continuations from identical near-critical states.
 All replicated particles, therefore, share the same past evolution up to the level-crossing time.  
The splitting procedure explores multiple possible future evolutions conditioned on having reached the given level, without resimulating past dynamics.

\subsubsection{Numerical results}

\begin{figure}
    \centering
    \includegraphics[width=0.95\linewidth]{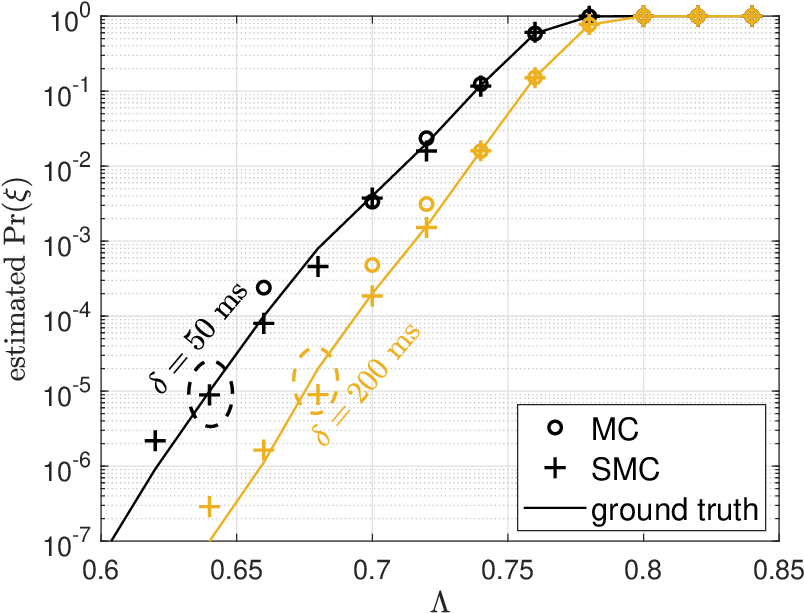}
    \caption{Estimated non-recovery probability $\Pr(\xi)$ versus the offered traffic rate $\Lambda$ for $\delta\in\{50,200\}$ ms and using MC and SMC simulation approaches. 
    }
    \label{fig:assessment}
\end{figure}

We illustrate some results in Fig.~\ref{fig:assessment} using the parameter values listed in Table~\ref{tab:baseline_params}, which are the defaults hereafter unless otherwise specified. We show the estimated probability of the resilience failure event as a function of the offered traffic rate for both MC and SMC simulation frameworks under the same computational cost. For this, we assume no active or adaptive mitigation so that the control parameter $u_p$ 
is held constant. 
The results confirm the superiority of SMC over MC in the region of extreme/rare failure events, i.e., $\Pr(\xi)\le 10^{-3}$. Indeed, MC can only simulate $C_\text{mc}/J=5\times 10^{6}/(60\ \text{s}/50\ \text{ms})\approx 4167$ trajectories, limiting its fault estimate to values greater than $1/4167=2.4\times 10^{-4}$. For SMC, an exact analysis is cumbersome, but it is clear from Fig.~\ref{fig:assessment} how it can produce estimates for several orders of lower fault probabilities.

\subsection{Service Reconfiguration}
Now, we showcase SMC for fault management as discussed in Section~\ref{sec:reconf}. For simplicity, we consider policies $u$ actionable only after entering $\mathcal{L}_2$.
Consider also a finite set of candidate recovery policies $\mathcal{U}=\{u_0,\cdots,u_{|\mathcal{U}|-1}\}$, where each $u_i$ leads to a distinct recovery rate 
\begin{align}
    \nu(u_i)=\nu_0(1+i\rho'),\ \rho'\in(0,1],
\end{align}
with $\nu_0$ denoting the baseline recovery rate used prior to any reconfiguration and corresponding to $u_0$, and such that $\nu(u_{|\mathcal{U}|-1})\Delta\le 1$. 
We adopt the myopic policy selection rule introduced in Section~\ref{sec:reconf}, such that \eqref{eq:LJ} becomes
%
    $\mathcal{J}(u_i|x_2^\star)=\ln \hat{p}_2(u_i) + c_2(u_i)$. 
Also, we model the policy cost proportional to the relative acceleration of recovery, as
\begin{align}
    c_2(u_i)=\kappa (\nu(u_i)-\nu_0)/\nu_0,\label{eq:cost}
\end{align}
where $\kappa\ge 0$ is a scaling parameter regulating the trade-offs between both terms of $\mathcal{J}(u_i|x_2^\star)$.

\subsubsection{SMC-assisted reconfiguration}
$\hat{p}_2(u_i)$ is estimated via simulation by exploiting the SMC checkpoints at level $\ell_2$. Specifically, upon first reaching $\ell_2$, the corresponding checkpoint state $x_2^\star$ is used as a common starting point, from which $N'$ independent trajectory continuations are generated under policy $u_i$, each with fresh realizations of the stochastic stress and recovery dynamics. Then, the policy that provides the smallest $\mathcal{J}(\cdot)$ is selected.
Simulations for the estimation of the resilience fault probability are continued then from $\ell_2$ onward with new random processes and the selected policy in place until the end of the horizon.

\begin{figure}
    \centering
    \includegraphics[width=0.95\linewidth]{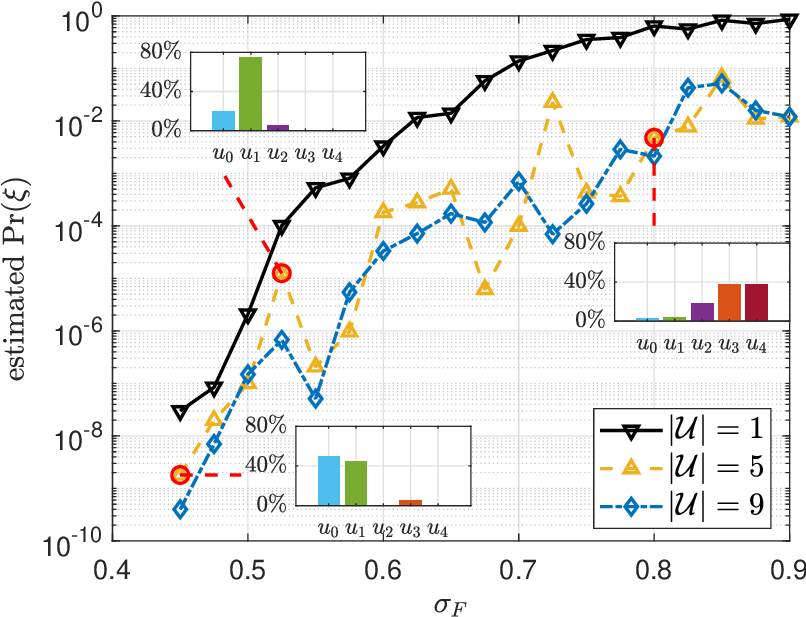}
    \caption{Estimated non-recovery probability $\Pr(\xi)$ versus the standard deviation of the latent stress $\sigma_F$ with the proposed SMC-assisted reconfiguration framework and policy sets $\mathcal{U}$ of different dimensions. 
    We set $\rho'=1/2$, $\kappa=1/2$, and $N'=25$. For the case of $|\mathcal{U}|=5$, we also plot the relative selection frequency of each policy for $\sigma_F\in\{0.45, 0.575, 0.8\}$.}
    \label{fig:reconf}
\end{figure}

\subsubsection{Numerical results}
Fig.~\ref{fig:reconf} illustrates the behavior of the estimated resilience fault probability and policy selection as a function of the latent stress variability across different dimensions of the policy sets under the proposed SMC-assisted reconfiguration framework. For this, we consider no policy reconfiguration, i.e., $|\mathcal{U}|=1$, as in the case of Section~\ref{sec:resR}, and also some reconfigurability using $|\mathcal{U}|=5$ and $|\mathcal{U}|=9$. 
We can corroborate that more comprehensive policy sets, i.e., larger $|\mathcal{U}|$, allow for more flexible recovery-cost trade-offs, often leading to improved recovery while still accounting for the policy costs \eqref{eq:cost}. Notably, under low stress, i.e., small $\sigma_F$, lower-rank policies are preferred due to their lower cost, while higher-rank, hence more disruptive, policies gain more traction as stress increases. 

%

\section{Conclusions and Outlook}\label{sec:conclude}
This work presented SMC for resilience assessment and control in networked systems. By formulating resilience failure events as path-dependent conditions over staged system dynamics, we showed how SMC enables efficient estimation of rare non-recovery probabilities through a multilevel splitting approach. The use of fixed, semantically meaningful levels allows for an interpretable decomposition of resilience into successive phases, while the proposed budget-adaptive population control mechanism ensures that computational effort is dynamically concentrated on the most critical transitions under a fixed total computation cost. The framework was also extended beyond estimation to support mitigation and control. This leverages SMC's checkpointing structure for policy evaluation and comparison at intermediate resilience levels, and for state-contingent policy selection via simulation-based lookahead. Overall, our proposed framework provides a unified approach to both assessing resilience and guiding recovery actions based on their impact on future system evolution, and its effectiveness was illustrated through a delay-critical wireless network use case. 

Future work will focus on assessing the potential of the proposed framework for more complex system dynamics and distributed control mechanisms. 
Also, beyond policy evaluation, we will consider the proposed framework as a data-generation mechanism for learning-based resilience control by exposing near-critical and post-fault states that are effectively unobservable under nominal sampling. 
Another relevant direction is the integration of data-driven models to approximate or accelerate the underlying simulator, enabling scalable resilience analysis in large-scale 6G systems.

\bibliographystyle{ieeetr}
\bibliography{ref}

\end{document}